# Electrohydrodynamic interaction between droplet pairs in confined shear flow

Somnath Santra[1], Diptendu Sen[2], Sayan Das[1], Suman Chakraborty[1],†

[1]Department of Mechanical Engineering, Indian Institute of Technology Kharagpur,
Kharagpur – 721302, India
[2]Department of Mechanical Engineering, Jadavpur University, Kolkata-700032, India

The present study deals with the numerical as well as asymptotic analysis of the electrohydrodynamic interaction between two deformable droplets in a confined shear flow. Considering both the phases as leaky dielectric, we have performed numerical simulations to study the effect of channel confinement on the drop trajectories in the presence of a uniform electric field. Two important varieties of motion are identified in the present analysis, namely (i) the reversing motion and (ii) the passing over motion. The study suggests that conversion of the passing over motion to the reversing motion or vice versa is possible via modulating the strength of the imposed electric field. Such a conversion of the pattern of droplet migration is also possible in a confined domain due to change in different electrical properties of the system (for instance conductivity). The present numerical model is also able to predict the pattern of the trajectory of individual droplets depending on the initial distance separating the two. For example, a smaller initial distance results in a reversing motion whereas a passing over motion is predicted when the distance between the droplets is significantly large. However, the final positions of the droplets are found to be independent of their initial positions. Interestingly, presence of electric field is found to prevent droplet coalescence to a certain extent depending on its strength, thus rendering the emulsion stable. A small deformation asymptotic model is also developed under the assumption of negligible fluid inertia to support the numerical results for the limiting case of an unbounded flow. The current investigation successfully presents a novel technique to predict the precise positions of a system of droplets in a micro-channel and how electric field can be used as a tool to modulate droplet trajectories in an emulsion. Such a study has a wide potential towards application in design and functionalities of several modern-day droplet based micro fluidics devices.

† E-mail address for correspondence: suman@mech.iitkgp.ernet.in

# 1. Introduction

The area of microfluidics has emerged in the last two decades as a promising interdisciplinary technology and has drawn significant attention of different research communities since then. It involves working with volume of fluids in the range of microliters to picoliters and has a number of applications. Among them, the deformation as well as motion of a liquid droplet in the presence of a background flow is of particular interest due to its wide ranging applications in emulsification (Lobo and Svereika, 2003; Jafari *et al.*, 2008), polymer blending (Ionescu-Zanetti *et al.*, 2004; Ho, Kim and Weitz, 2008), recovery of oils (Nilsson *et al.*, 2013; Lifton, 2016) and drug delivery (Kleinstreuer, Li and Koo, 2008; Xu *et al.*, 2009) to name only a few. Growing interest is developing in this field due to its advantages such as rapid mass delivery and heat transfer, low sample consumption and waste generation. In polymer blending, the microscopic morphology of a polymer blend plays a prime role in determining material properties of industrial interest, such as mechanical strength and permeability and it is a well known fact that the morphology development under flow is mainly governed by the two key phenomenon: break up and coalescence of the droplets forming the dispersed phase. However, precise control over the size and morphology of the droplets has always been a challenge. So, for a better control over the relevant technological processes, it is important to know the underlying hydrodynamics of these processes as well as to identify different actuating forces that can control the droplets motion. Common factors influencing droplet dynamics include the fluid inertia (Ho and Leal, 1974; Mortazavi and Tryggvason, 2000; Chen *et al.*, 2014), surface deformation (Goldsmith and Mason, 1962; Chaffey, Brenner and Mason, 1965; Haber and Hetsroni, 1971; Wohl and Rubinow, 1974; Stan *et al.*, 2011; Mandal, Bandopadhyay and Chakraborty, 2015), presence of surfactants (Hanna and Vlahovska, 2010; Pak, Feng and Stone, 2014; Das *et al.*, 2017; Das, Mandal and Chakraborty, 2017a), magnetic fields (Zhang *et al.*, 2009), acoustic waves (Franke *et al.*, 2009) and thermo capillary stresses (Baroud *et al.*, 2007; Das, Mandal and Chakraborty, 2017b). Apart from these, externally applied electric fields (Hase, Watanabe and Yoshikawa, 2006; Ristenpart *et al.*, 2009; Zagnoni and Cooper, 2009; Kurimura *et al.*, 2013; Mhatre and Thaokar, 2013; Vajdi Hokmabad *et al.*, 2014) can also act as a means for controlling droplet motion.

When a liquid droplet is kept in another fluid with different dielectric properties, the difference leads to a build-up of electrical stresses, causing the droplet to deform. Until Taylor's work, fluids were supposed to be entirely conducting or pure dielectrics. According to electrostatics, if the fluids are treated as perfect dielectrics, then the electrical stresses act only normal to the interface. So, an isolated droplet should deform into a prolate shape with no fluid motion in the final equilibrium state. However, experiments of Allan and Mason (Allan and Mason, 1962) showed oblate deformation. This led to Taylor giving the leaky dielectric theory in 1966 (Taylor, 1966). According to this theory, fluids should not be considered as perfect conductors or dielectrics, rather they are supposed to have finite permittivity and conductivity to allow for the accumulation of free charges at the interface. Electric field, acting on this charge, will cause imbalance in both normal and tangential stresses, making oblate deformation possible. Also, the imbalance in shear stress at equilibrium has to be balanced by hydrodynamic shear stresses, resulting in fluid motion both inside and outside the droplets. To specify the nature of deformation, Taylor introduced a deformation characteristic function analytically. A negative value of the function stands for a oblate deformation, while prolate deformation is characterized by a positive magnitude of the function. After the pioneering work of Taylor, several analytical, numerical and experimental studies (Baygents, Rivette and Stone, 1998; Chabert, Dorfman and Viovy, 2005; Hase, Watanabe and Yoshikawa, 2006; Link *et al.*, 2006; Mandal, Bandopadhyay and Chakraborty, 2016; Bandopadhyay, Mandal and Kishore, 2017) involving the deformation and motion of liquid droplets in presence of electric field have been done. Apart from this, several efforts has been made to show the non-trivial interaction of the droplet pairs in presence of electric field. In a related study, Mohammadi et al. (Mohammadi, Shahhosseini and Bayat, 2014) have made a numerical study of the collision and coalescence of water droplets in an electricfield. They found a relation between the electric field magnitude, droplet size, and initial separation distance. Later on, Luo et al. (Luo, Schiffbauer and Luo, 2016) have studied the effect of non uniformity of electric field on droplet coalescence and showed that droplets moved collectively towards the converging field due to the positive dielectrophoresis unlike uniform electric field where no net droplet movement was observed. Besides these, the deformation and interaction of a droplet pair under applied ac electric field and the dynamics of the liquid bridge and their tip streaming(Chen *et al.*, 2015), electrostatic interactions between two charged droplets (Su, 2006), the oscillatory coalescence of droplets under an ac electric field (Choi and Saveliev, 2017) are also worth mentioning.

Numerous studies have also been done on the interaction of droplets in presence of shear flow. In a pioneering work, Guido and Simeone (Guido and Simeone, 1998) showed the collision of two droplets under a simple shear flow with the help of computer assisted microscopy. Later on, Olapade et al. (Olapade, Singh and Sarkar, 2009) demonstrated the droplet trajectory in free shear for finite inertia for two deformable droplets. Lac et. al (Lac, Morel and Barthès-Biesel, 2007; Lac and Barthès-Biesel, 2008) showed the nature of hydrodynamic interaction between two capsules in a confined shear flow. The effect of inertia on the nature of the above interaction was studied by Doddi and Bagchi (Doddi and Bagchi, 2008). After the literature review, we have found that the interaction between droplet pairs in combined presence of uniform electric field and shear flow has not been explored much till date. In addition to this, the effect of domain confinement on droplets interaction has also not got much attention though it has considerable impact in droplet deformation characteristics (Sibillo *et al.*, 2006; Vananroye, Puyvelde and Moldenaers, 2006; Vananroye, Van Puyvelde and Moldenaers, 2007; Vananroye *et al.*, 2010; Barai and Mandal, 2016). In a recent study, Chen and Wang (Chen and Wang, 2014) showed the effect of domain confinement on the motion of droplet pairs in confined shear flow. In the present analysis, we have performed a numerical analysis to explore the underlying hydrodynamics behind the non-trivial interaction of the deformable droplets in combined presence of transverse electric field and background shear flow in micro-confined environment.

The objectives of our study are: (i) effect of electric field strength on the pair wise motion of the droplets, (ii) effect of domain confinement on the pattern of droplet motion (iii) effect of electric properties of the system on their pair wise motion and (iv) effect of initial lateral separation distance on the trajectory of the droplet pairs (v) presence of electric field on the coalescence phenomenon of droplet pair in unconfined and confined medium.

## 2. Numerical model setup

In our present study, we have taken two leaky dielectric droplets (radius $a$, density $\rho$, viscosity $\mu_i$, electrical permittivity $\varepsilon_i$ and conductivity $\sigma_i$) suspended in another leaky dielectric liquid medium (of density $\rho$, viscosity $\mu_o$, electrical permittivity $\varepsilon_o$ and conductivity $\sigma_o$) subjected to background shear flow as well as uniform transverse electric field, $E$ applied between the top and bottom walls as shown in figure 1. The upper plate is moving with a velocity $U$ towards the positive x direction while the lower plate traverses with the same velocity in the opposite direction. The plates are located at a distance $H$ above and below the y axis. The droplets are separated by a distance $\Delta x$ in the horizontal direction and $\Delta y$ in the lateral direction. The electric potentials of the lower and upper walls are $\phi = -E_0 H$ and $\phi = E_0 H$ respectively. For the present analysis, we have used Cartesian coordinate system attached at the centerline of the parallel plate channel.

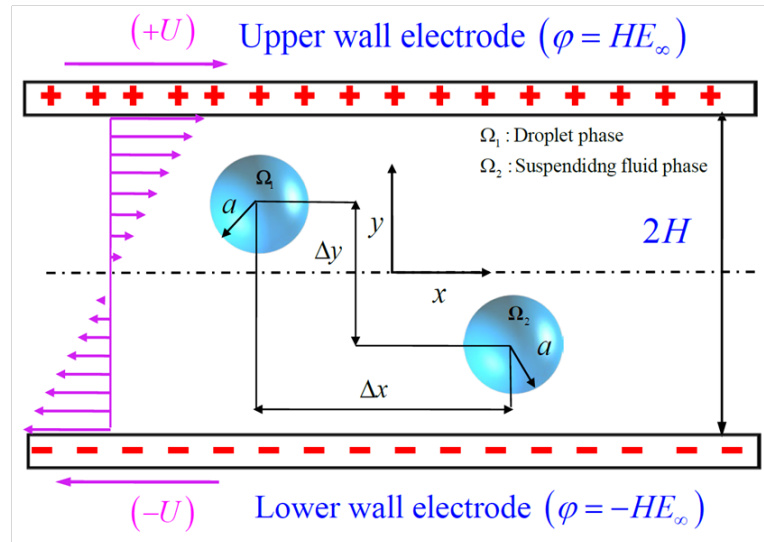

FIGURE 1. Schematic of a physical system depicting the position and configuration of two liquid droplet of radius $a$ suspended in a confined shear flow under the influence of uniform electric field. The top wall-electrode and bottom wall-electrode have electric potential $HE_\infty$ and $–HE$ respectively. $2H$ is the distance between the walls and $E_\infty$ is the magnitude of applied electric field.

## 2.1. Numerical simulation- phase field formalism

In the present analysis, we have used phase field formalism (Jacqmin, 1999; Badalassi, Ceniceros and Banerjee, 2003; Mandal, Ghosh and Bandopadhyay, 2017) for assessing the essential dynamics of two fluid system. Several researchers (Wang, Qian and Sheng, 2008; Chaudhury, Mandal and Chakraborty, 2016; Mandal, Ghosh and Bandopadhyay, 2017) have reported this formalism in their studies and mentioned that it is fit for handling the diffuse interface of a binary fluid system with good accuracy. This approach involves a phase field variable $\varphi$ for determining the distribution of the constituting fluid element, which varies from -1 in one phase to 1 in the other phase. For our case we have taken $\varphi$ as -1 inside the droplets and 1 in the carrier fluid. The interfacial zone is denoted by a diffuse region where $\varphi$ varies in between -1 and 1.

The Ginzburg-Landau free energy for a Newtonian two fluid system in terms of phasefield variable is expressed as

$$F = \int_{\forall} \left\{ f(\varphi) + \frac{1}{2} \gamma \xi |\nabla \varphi|^2 \right\} d\forall \qquad (1)$$

The first term on the right side denotes the bulk free energy( which represents the configuration between the two fluids at equilibrium condition ), while the latter term is for the excess free energy at the interface. $\gamma$ stands for the surface tension and $\xi$ symbolized the characteristic length scale for the diffuse interface thickness. The expression for the bulk energy is provided as (Jacqmin, 1999; Mandal, Ghosh and Bandopadhyay, 2017)

$$f(\varphi) = \frac{\gamma}{4\xi}(1-\varphi^2)^2 \qquad (2)$$

One must acknowledge that the chemical potential and total energy of the system are related with each other in the following way,

$$G = \frac{\delta F}{\delta \varphi} = -\gamma \xi \nabla^2 \varphi + \frac{\partial f}{\partial \varphi} \qquad (3)$$

The value of *G*, equated to zero, gives us the equilibrium interface profile between the two phases.

## 2.2. Governing equations and boundary conditions

To find out the value of the order parameter $\varphi$, we make use of the Cahn Hilliard equation. This equation, consisting of diffusion as well as advection of the order parameter can be represented as

$$\frac{\partial \varphi}{\partial t} + \mathbf{u}.\nabla \varphi = \nabla \cdot \left( M_\varphi \nabla G \right). \tag{4}$$

Here, $M_f$ and $G$ stand for the interface mobility factor and the chemical potential respectively. According to the phase field model, the fluid properties are defined as

$$\left. \begin{array}{l} \bar{\rho} = \dfrac{(1+\varphi)}{2}\rho_i + \dfrac{(1-\varphi)}{2}\rho_0, \quad \bar{\mu} = \dfrac{(1+\varphi)}{2}\mu_i + \dfrac{(1-\varphi)}{2}\mu_0 \\[6pt] \bar{\varepsilon} = \dfrac{(1+\varphi)}{2}\varepsilon_i + \dfrac{(1-\varphi)}{2}\varepsilon_0, \quad \bar{\sigma} = \dfrac{(1+\varphi)}{2}\sigma_i + \dfrac{(1-\varphi)}{2}\sigma_0 \end{array} \right\} \tag{5}$$

In the numerical simulation, we have considered that density and viscosity of both the fluids are equal. For the absence of magnetic field and negligible dynamic current, the electric field equation is interpreted as (Reddy and Esmaeeli, 2009)

$$\bar{\nabla} \cdot \left( \bar{\sigma} \mathbf{E} \right) = 0 \tag{6}$$

Due to the irrotationality of the electric field, it is related with the electric potential as $\mathbf{E} = -\nabla \phi$ and the Equation(6) reduces to

$$\bar{\nabla} \cdot \left( \bar{\sigma} \bar{\nabla} \phi \right) = 0 \tag{7}$$

Electric potential is specified at the both wall as

$$\left. \begin{array}{l} \text{top wall: } \phi = HE_\infty \quad \text{at } \bar{y} = H, \\ \text{bottom wall: } \phi = -HE_\infty \quad \text{at } \bar{y} = -H, \end{array} \right\} \tag{8}$$

For obtaining the velocity field and pressure field, the continuity equation and Navier-Stokes have been used in the following form

$$\nabla \cdot \bar{\mathbf{u}} = 0, \; \rho \left( \frac{\partial \bar{\mathbf{u}}}{\partial t} + \bar{\nabla} \cdot (\bar{\mathbf{u}}\bar{\mathbf{u}}) \right) = -\bar{\nabla} \bar{p} + \bar{\nabla} \cdot \left[ \left\{ \mu(\bar{\nabla}\bar{\mathbf{u}} + (\bar{\nabla}\bar{\mathbf{u}})^T) \right\} \right] + \bar{G}\bar{\nabla}\varphi + \bar{\mathbf{F}}^E \tag{9}$$

The interfacial tension force has been taken into consideration by the term $\bar{G}\bar{\nabla}\varphi$ and $\bar{\mathbf{F}}^E$ denotes the volumetric electrical force at the interface of the droplet and expressed as $\mathbf{\bar{F}}^{E} = \bar{\nabla} \cdot (\bar{\varepsilon} \bar{\nabla} \bar{\phi}) \bar{\nabla} \bar{\phi} - |\bar{\nabla} \bar{\phi}|^{2} \bar{\nabla} \bar{\varepsilon}/2$. The velocity field satisfies the no-slip and no penetration boundary conditions at the two walls ($\mathbf{n_s}$ is the unit normal vector at the solid surface)

$$\begin{aligned} \text{top wall:} \bar{\mathbf{u}} \cdot \mathbf{n}_s = 0, \bar{\mathbf{u}} - (\bar{\mathbf{u}} \cdot \mathbf{n}_s)\mathbf{n}_s = +U \quad \text{at } \bar{y} = H, \\ \text{bottom wall:} \bar{\mathbf{u}} \cdot \mathbf{n}_s = 0, \bar{\mathbf{u}} - (\bar{\mathbf{u}} \cdot \mathbf{n}_s)\mathbf{n}_s = -U \quad \text{at } \bar{y} = -H, \end{aligned} \quad (10)$$

For non-dimensionalizing the governing equations and boundary conditions, we have used appropriate caharcteristic scale: Length is non-dimensionalized by $a$, velocity by $S_R a$ (Where $S_R$ is the shear rate), electric field by $E_\infty$, viscous stress by $\mu_e u_s/a$, and electric stress by $\varepsilon_e E_\infty^2$. After the non-dimensionalization, we have fixed some non-dimensional numbers and material property ratio: capillary number $Ca = \mu_e u_s/\gamma$ (ratio of viscous to capillary stresses), Masson number $M = \varepsilon_e E_\infty^2 a/\mu_e u_s$ (ratio of electric to viscous stresses), $S = \varepsilon_i/\varepsilon_i$, Reynolds number $Re = \rho G a^2/\mu_e$ (which signifies the relative strength of inertial stress as compared with viscous stress), viscosity ratio $\lambda = \mu_i/\mu_e$, conductivity ratio $R = \sigma_i/\sigma_e$, and permittivity ratio $S = \varepsilon_i/\varepsilon_e$. The non-dimensional forms (the over bar sign has been removed) of the governing equations are represented below

$$\nabla \cdot (\sigma \mathbf{E}) = 0 \quad (11)$$

$$\nabla \cdot \mathbf{u} = 0 \quad (12)$$

$$\frac{\partial \varphi}{\partial t} + \mathbf{u} \cdot \nabla \varphi = \frac{1}{Pe} \nabla \cdot M_f (\nabla G); \text{ where } G = \frac{1}{Cn}(\varphi^3 - \varphi) - Cn \nabla^2 \varphi \quad (13)$$

$$Re\left(\frac{\partial \mathbf{u}}{\partial t} + \nabla \cdot (\mathbf{u}.\mathbf{u})\right) = -\nabla p + \nabla \cdot \left[\left\{\nabla \mathbf{u} + (\nabla \mathbf{u})^T\right\}\right] + \frac{1}{Ca} G \nabla \varphi + M \mathbf{F}^E \quad (14)$$

The non dimensional fluid properties are represented as

$$\begin{aligned} \rho = \frac{(1+\varphi)}{2}\rho_r + \frac{(1-\varphi)}{2}, \quad \mu = \frac{(1+\varphi)}{2}\lambda + \frac{(1-\varphi)}{2} \\ \varepsilon = \frac{(1+\varphi)}{2}S + \frac{(1-\varphi)}{2}, \quad \sigma = \frac{(1+\varphi)}{2}R + \frac{(1-\varphi)}{2} \end{aligned} \quad (15)$$

It is also important to mention that the velocity field, pressure field and the phase field variable are periodic in the horizontal direction and expressed as

$$\left.\begin{array}{ll}\text{(i)} & \mathbf{u}(\mathbf{x}) = \mathbf{u}(\mathbf{x}+L), \\ \text{(ii)} & p(\mathbf{x}) = p(\mathbf{x}+L), \\ \text{(iii)} & \varphi(\mathbf{x}) = \varphi(\mathbf{x}+L).\end{array}\right\} \quad (16)$$

## 3. Results and Discussion:

### 3.1 Effect of electric capillary number on the droplet trajectory

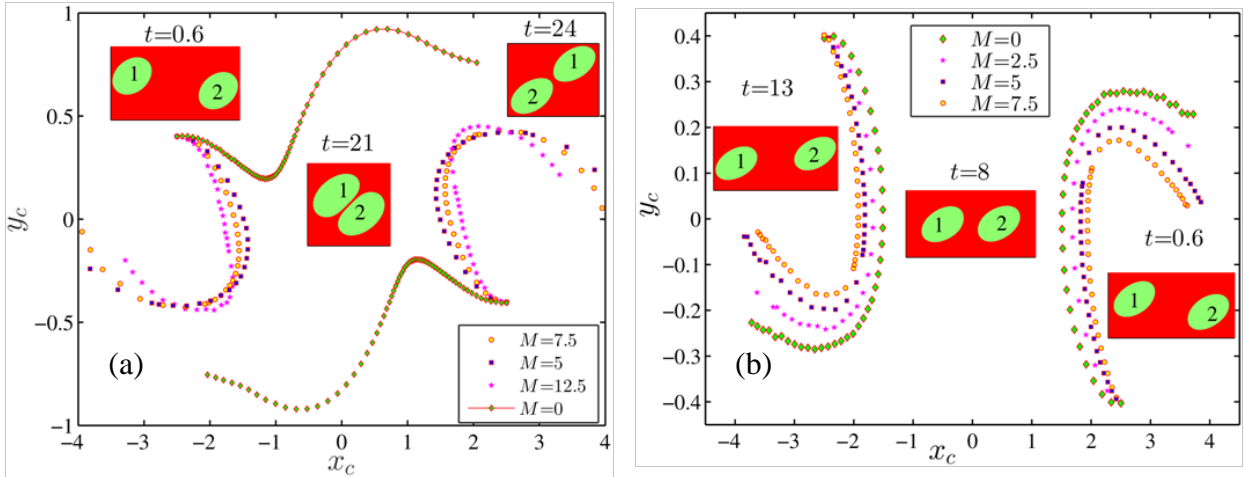

FIGURE.2 Variation of droplet trajectory for a system having $R=2$, $S=0.5$, $Ca=0.2$ for (a) $Wc=0.20$ (b) $Wc=0.40$. The contour plots in the insets shows the particular case, where there is no presence of any electric field ($M = 0$). Others parameters are: $\lambda= 1$, $Re = 0.2$

Figure 2 shows the effect of electric capillary number on the droplet trajectory of a LD-LD system having $R>S$. In unbounded domain, we observe that with increasing electric capillary number, the passing over motion (observed when there is no electric field) changes to a reversing motion of gradually decreasing radius of curvature. Whereas for a confined domain, the radius of curvature simply decreases with increasing electric field but the same reversing motion is retained for each case.

The reason behind the observed phenomenon is now provided. When interacting droplet pairs are subjected to linear shear flow, they initially moves towards the centreline due to back ground flow and when they come closer to each other, under goes into two type of motion (i) passing over motion and (ii) reversing motion. In passing over motion, passing over streamlines pass from the upper and lower half of the domain that generates drag force. This drag force is responsible for such type of motion. Similarly, reversing motion characterized by the reversing flow that is represented by the reversing stream line. Due to the application of electric field, one additional elctrohydrodynamic force ($F_{EHD}$) is created that can direct the droplet towards centreline or away from the centreline depending on the direction of fluid flow circulation. For $R>S$, the direction of fluid flow circulation takes place from equator to pole and $F_{EHD}$ direct the droplet toward centreline and a reverse scenario is observed for $S>R$, where the fluid flow circulation takes place from pole to equator. Now, for an unbounded domain without any electric field, the droplets are exposed more to the passing over streamline region than the reversing ones, leading to the passing over motion. However, in presence of the electric field, the $F_{EHD}$ acts on the droplet and the droplets show a reversing motion. Since the magnitude of this $F_{EHD}$ is proportional to the electric field, they move towards the centreline with smaller radii of curvature at higher electric field strength.

Now, for a confined domain, along with the usual effects, the electric potential gets amplified due to wall confinement that increases the strength of $F_{EHD}$. Furthermore, a repulsive force ($F_R$) is also acting due to wall confinement that also tries to drive the droplet towards the centreline. More over in confined domain, reversing flow regime in the fluid domain gets bigger that increases the role reversing flow. So, even when there is no electric field in this case, the droplets are exposed more to the reversing flow region (due to wall confinement) than the passing over region and follow a reversing motion. Due to the application of electric field, $F_{EHD}$ is generated that results in the normal reversing flow for a system having $R>S$. With increasing electric field, the magnitude of the $F_{EHD}$ on the droplets increases, resulting in the smaller radii of curvature of their motion.

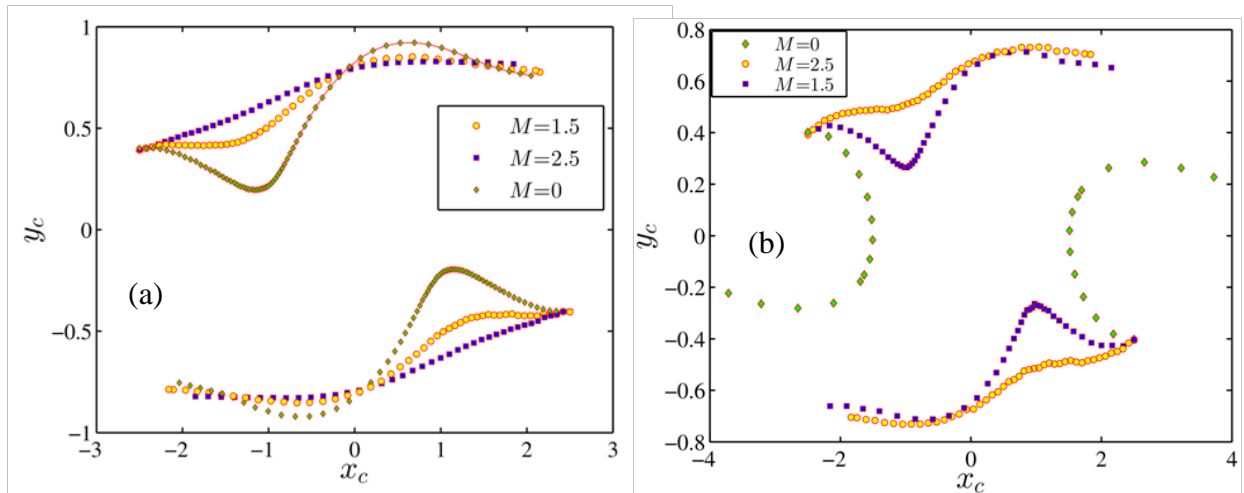

Figure 3 Variation of droplet trajectory for a LD-LD system having $R=0.5$, $S=2$, $Ca=0.2$ for (a) $Wc = 0.20$ (b) $Wc=0.40$. Others parameters are: $\lambda= 1$, $Re = 0.2$

Figure 3 shows the variation of droplet trajectory for a LD-LD system having $S>R$. In this case, we observe that the passing over nature remains intact but the droplet tend to repel each other more with increasing strength of electric field for an unbounded domain. For a confined domain, the reversing motion observed without any electric field, changes to a passing over motion with increasing mutual separation as the electric field strength increases.

For a system having $S>R$, the electro hydrodynamic force tends to move the droplet towards the wall while the other two forces (wall induced repulsion force and hydrodynamic force due to the background flow) act in the same way as discussed earlier. For an unbounded domain, the normal passing over motion is observed due to the droplet being more exposed to the passing over region. They show a tendency of coming towards each other due to the $F_{EHD}$. As the Mason number ($M$) increases, the magnitude of $F_{EHD}$ increases, hence the droplet tend to repel each other more, as is supported by the graph. For a confined domain, in absence of any electric field, the normal reversing motion is observed as explained for figure 1. However, in this case, the final motion is the combined effect of wall induced repulsive force and electro hydrodynamic force. Though $F_R$ tries to bring the droplets towards each other, the $F_{EHD}$ forces them to move towards the wall. At higher electric field strength, the strength of the $F_{EHD}$ is much higher that drives the droplet to the passing over region and the droplet undergoes into passing over motion.

Now, we have constructed a region diagram that shows two distinct regions of droplet motion in confined domain based on the values of ($Ca$, $Ca_E$) as shown in figure 4. The points with yellow-coloured triangular markers show the values of ($Ca$, $Ca_E$) where the droplet undergoes into reversing motion. Similarly, the points with blue-coloured triangular markers represent the values of ($Ca$, $Ca_E$), at which the droplet undergoes passing over motion. From the regime diagram, it has been obtained that the movement of droplets changes from passing over to reversing motion with increase in the value $Ca$. At higher values of $Ca$, the deformation is more that exposes a large portion of the droplet into reverse flow region and the droplet follows reverse flow motion. However, at higher values of $Ca$, $F_{EHD}$ force tries to move the droplet towards the wall and large portion of the droplet falls into the passing over flow region that creates a passing over motion of the droplet.

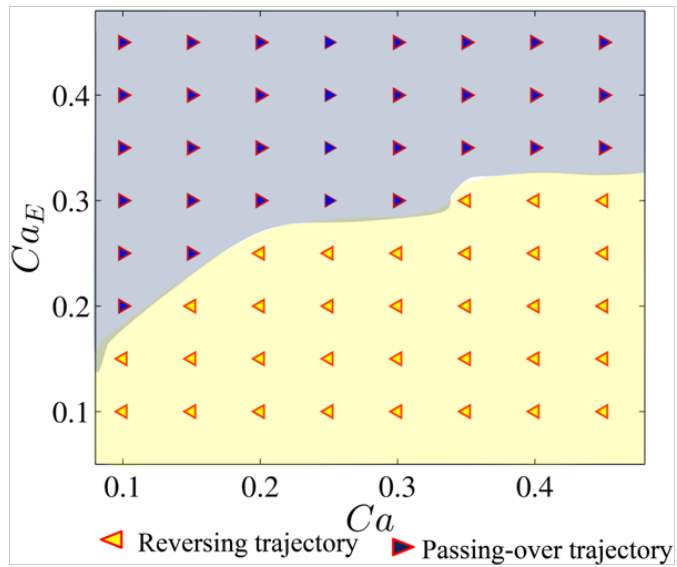

FIGURE 4 Regime diagram based on the value of ($Ca$, $Ca_E$). Others parameters are: $\lambda= 1$, $Re = 0.2$, $S=2$, $R=0.5$, $Wc =0.4$

3.2 Effect of domain confinement on droplet motion

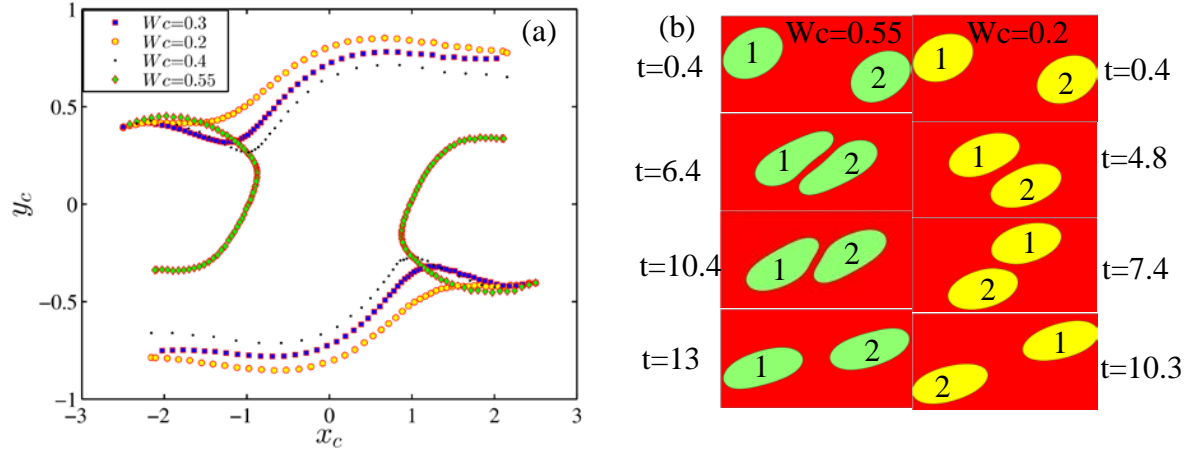

FIGURE 5 (a) Variation of droplet trajectory with domain confinement for a system having $S=2$, $R=0.5$, $M=1.5$, (b) Droplet contour at different times. Others parameters are: $\lambda= 1$, $Re = 0.2$

Figure 5 shows the variation of droplet trajectory for different confinement ratios for a LD-LD system having $S>R$. We observe that, for very small confinement ratio or unbounded domain, the usual passing over motion is observed. However, as the confinement increases, the droplets come closer to one another, and after a certain limiting value ($Wc\approx0.55$), this changes to a reversing motion.

The normal passing over motion is observed due to the reasons as explained earlier. Now, for a system having $R<S$, it is true that the electric field strength increases with increases domain confinement that enhances the magnitude of $F_{EHD}$. However, the magnitude of the repulsive force also increases with increase in domain confinement that compensates for the EHD force and tends to move the droplets into reverse flow region. Also with increase in confinement, the reversing streamline zone increases and the droplets are exposed more to the reversing zone than to the passing over zone, resulting in the aforementioned behavior.

### 3.3  Effect of conductivity ratio on the droplet trajectory

Figure 6 shows the variation of droplet trajectory in confined domain for different conductivity ratio. From the figure, it has been obtained that their nature of motion changes to a reversing motion from passing over motion with gradually decreasing radius of curvature as the value of $R$ increases.

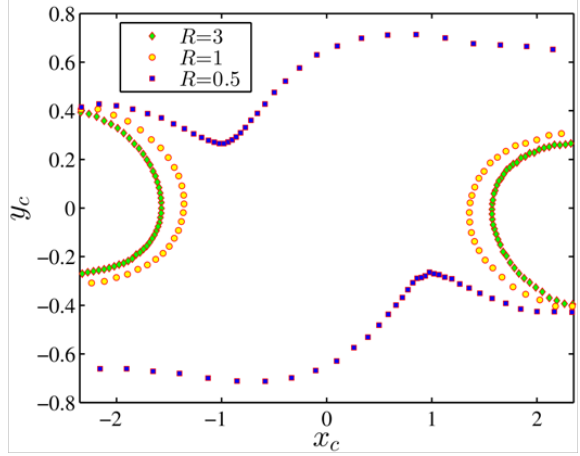

FIGURE 6 Variation of droplet trajectory for different conductivity ratio. Other parameters are $S$ =0.5, $\lambda$=1, $Wc$=0.4, $Re$=0.2, $Ca$=0.2, $M$=1.5

The physical reason behind the observed phenomenon is provided in the present section. When the value of $R$ is less, the deformation of the droplet is also less, which exposes a large portion of the droplet in passing-over flow regime. Furthermore, as the value of $R$ is less, the confinement induced electric field strength is also weak. Due to that reason, the magnitude of $F_{EHD}$ is also not too strong to move the droplet in to reverse flow regime and the droplet finally follows a passing over motion. For the considered values of ($R$, $S$), $F_{EHD}$ tries to move the droplet towards the centreline. However, at higher values of $R$, the deformation of the droplet is more, and a large portion of the droplet is subjected to reverse flow region. Moreover, the strength of confinement induced the imposed electric field is also high leading to an increase in the electro hydrodynamic force. Hence, the tendency of the droplet to move towards the centreline increases, and they show a reversing motion.

### 3.4 Effect of lateral separation distance on droplet trajectory

Figure 7 depicts the variation of the droplet motion due to change in its initial lateral separation distance. It is observed that when the separation is larger, the droplets show a passing-over motion and the terminal position is independent of its initial position. The passing-over motion transits to a reversing motion when the distance decreases. It is important to mention, for reversing motion also, the terminal position of the droplet does not depend on its initial position. When the lateral separation distance is large, more volume of the droplets are exposed to the passing over streamline than the reversing streamline. This results in the passing over motion. However, as they are brought closer, more volume of the droplets are exposed to the reversing zone, resulting in the transition to reversing motion.

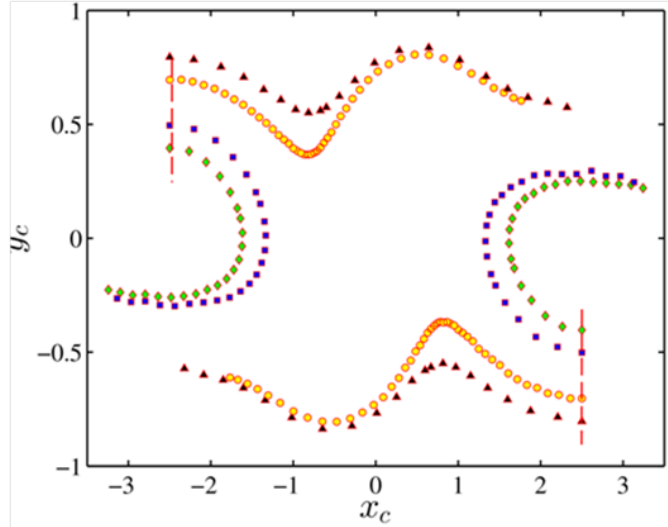

FIGURE.7 Variation of droplet trajectory with initial position (Δy (*t*=0)=0.8, 1, 1.4, 1.6) for a system having *R*=2, *S*=0.5.Others parameters are: λ= 1, *Re* = 0.2, *Wc*=0.4, *Ca*=0.2, *M*=1.5

3.5 Effect of electric field on the coalescence of the droplet pairs

Figure 8 shows the effect of electric field on the coalescence of a droplet pair of LD-LD system having *S>R* in unbounded domain. From the figure, it has been obtained that the droplet undergoes into coalescence in absence of electric field while in the presence of electric field, they show a passing over motion abolishing the tendency of coalescence.

The reason behind the observed phenomenon is now provided. In absence of electric field, in an unconfined domain, the hydrodynamic force due to back ground flow tends to bring the droplet towards the centreline. As the droplet come sufficiently close to each other, they form a rotating pair. Also, at the same time, the thin film between them is getting drained. When the film becomes sufficiently thin, the Van der Waals force between them causes the film to rupture, and they coalesce. However, when an electric field is applied, an additional $F_{EHD}$ comes into play. For the considered values of (*R*, *S*) this force tries to move the droplets towards the wall and a large portion of the droplet is exposed to the passing over flow regime that finally creates the passing over motion of the droplet.

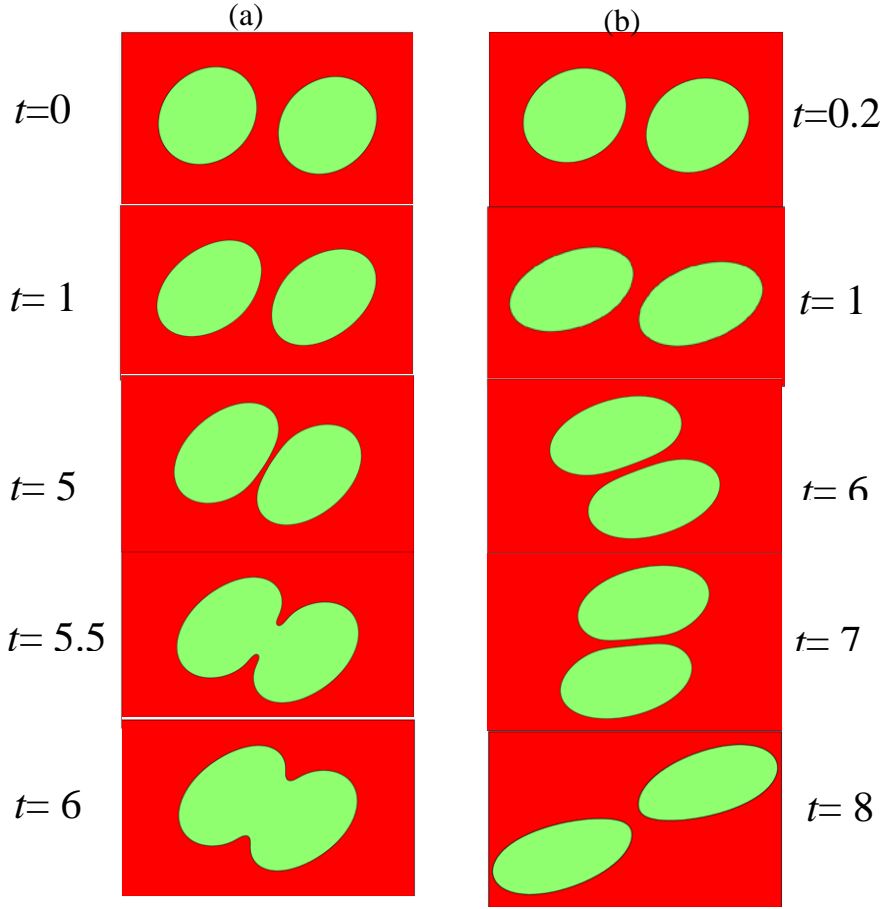

FIGURE 8 Droplet behavior in unbounded domain for (a)$M=0$ and (b)$M=1.2$. Others parameters are: $R=0.5$, $S=2$ $\lambda=1$, $Re=0.2$, $Wc=0.2$, $Ca=2$, $\Delta y(t=0)=0.2$, $\Delta x(t=0)=2.5$

Similarly, figure 9 shows the coalescence of a droplet pair of a LD-LD system having $R>S$ in confined domain. The figure shows that the droplets still coalesce in confined domain when there is no electric field. On applying an electric field, they show a reversing motion in contrast to the passing over motion in an unbounded domain. For the considered value of $(R, S)$, the deformation of the droplet is more in confined domain. So, a large portion of the droplets is exposed to reverse flow region. Furthermore, for the considered values of $(R, S)$, $F_{EHD}$ force tries to move the droplet towards the reverse flow region and its magnitude is also significantly high in confined droplet that creates a reversing motion of the droplet pair.

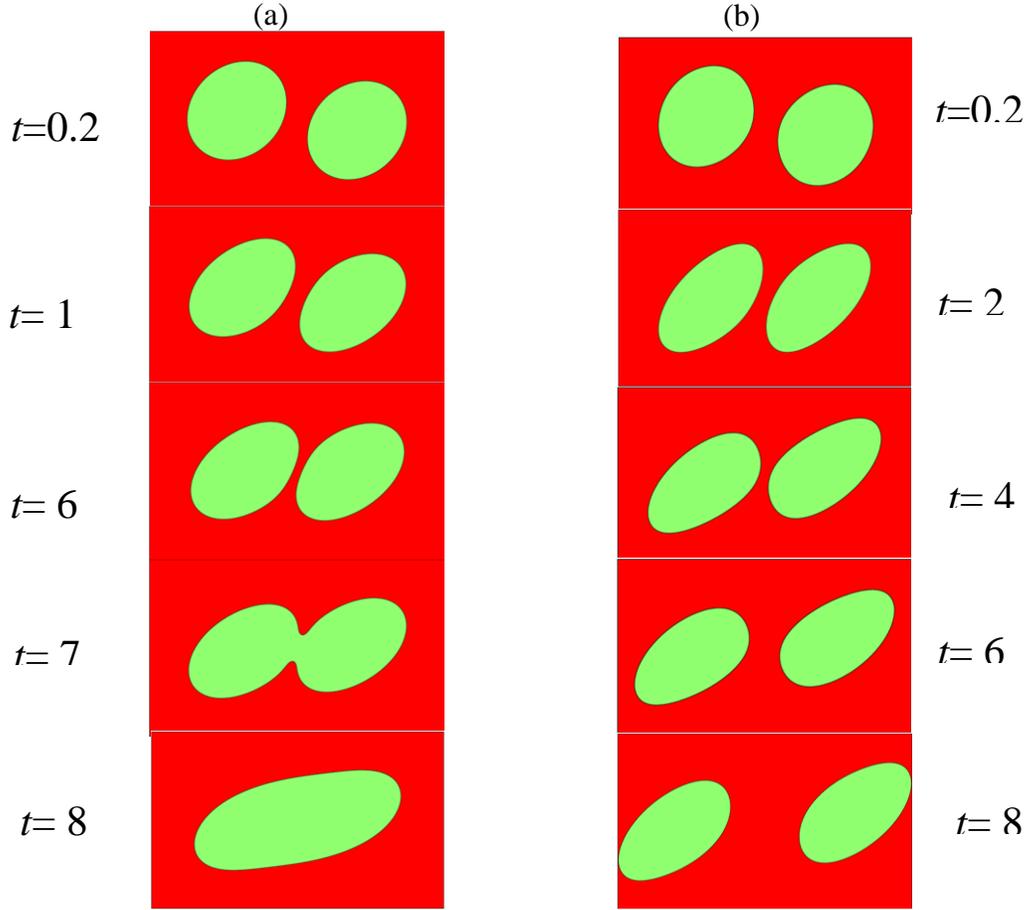

FIGURE 9 Droplet behavior in confined domain for (a) $M=0$ and (b) $M=5$. Others parameters are $\lambda = 1$, $Re = 0.2$, $Wc=0.2$, $Ca=0.2$, $\Delta y(t=0)=0.4$, $\Delta x(t=0)=2.52$

## 4. Conclusions

In the present analysis, we have performed numerical simulation to investigate the interaction of droplet pairs in a parallel plate channel under the combined presence of linear shear flow and uniform transverse electric field. In the present analysis, we have highlighted on the trajectory of droplet motion. The present study depicts that trajectory of droplet motion can be altered by tuning different parameters like confinement ratio, electric field strength and the electrical properties of the system. Through numerical simulation, we have also highlighted on the electro-coalescence of droplet both in confined and unbounded domain. After a detailed numerical analysis, we have reached to the following concluding points

1. For a LD-LD system having $R>S$, the trajectory of droplet motion changes from passing-over motion to reversing motion with increase in the value of $M$ in unbounded domain. On contrary, for a LD-LD system having $S>R$, the trajectory of the droplet motion changes from reversing motion to passing-over motion at higher values of $M$ in confined domain.

2. For a LD-LD system having *S*>*R,* with increase in the domain confinement, the droplet motion changes from passing over to reversing motion.

3. Pattern of droplet motion also changes from passing over to reversing motion with increase in the values of conductivity ratio of the system (*R*).

4. For a smaller lateral separation distance, the droplet pairs follow a reversing motion, whereas they follow passing-over motion if the lateral separation distance is large. The interesting fact is that for each type of motion, the terminal position is independent of its initial position.

5. Finally, this study reveals that the tendency of the electro-coalescence of the droplet is disappeared v due to the application of electric field both in confined and unbounded domain.

**ACKNOWLEDGMENTS**

S.S. is grateful to Dr. Shubhadeep Mandal for insightful discussions on droplet EHD**.**

**Appendix A: Validation of our numerical setup**

For checking the accuracy of the present numerical result, we have validated our result with three previously published works as shown in figure 10(a) and 10(b). First of all, we have made a comparison between trajectory of droplet pairs obtained from our numerical result with the result of Chen and Wang (Chen and Wang, 2014) both in confined and unconfined domain in absence of electric field and we have found a very good agreement between them.

For checking the applicability of present 2D numerical simulation, we have compared our simulation results with the existing experimental results for the deformation (*D*) of the

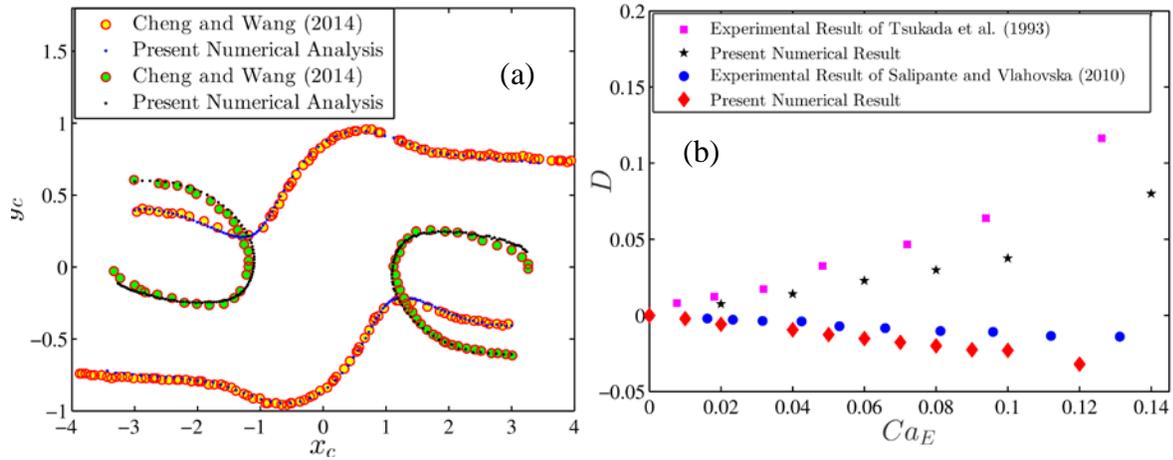

FIGURE 10 (a) Comparison between present numerical result and numerical result of Chen and Wang (Chen and Wang, 2014). Other parameters are $Ca$=0.2, $Re$=0.2, $\lambda$=1. (b) Comparison of Taylor deformation parameter between our numerical result and experimental results. The properties of fluids considered in the study of Tsukada et al. (Tsukada *et al.*, 1993) and Salipante and Vlahovska (Salipante and Vlahovska, 2010) are ($S$, $R$)=(1.72, 62.4), $\lambda$=1.47 and ($S$, $R$)=(0.57, 0.027), $\lambda$=1.47 respectively droplet in unbounded domain in presence of electric field. For a LD-LD system having $R$>$S$, we have compared the value of deformation parameter obtained from our result with result of Tsukada et al. (Tsukada *et al.*, 1993) and for $S$>$R$ , we have compared it with the result of Salipante and Vlahovska (Salipante and Vlahovska, 2010). In both the cases, we have found a very good qualitative matching.

**Appendix B: grid independency check**

For ensuring the accuracy of the present numerical solution, it is required to perform the grid independence and Cahn number independence studies. As in the present case, the grid size and $Cn$ number are same near the interface. So, a correct grid independence necessarily means a correct Cahn number independence study and vice versa.(Mandal *et al.*, 2015). For the $Cn$ independence test, we have evaluated the trajectory of droplet pair under transverse electric field for three different Cahn number ($Cn$=0.015, 0.020, 0.025) as shown in figure 11. The figure shows a negligible variation of droplet trajectories for the chosen $Cn$ numbers. Finally, we have chosen $Cn$=0.015 for the present analysis and all the plots in the present study have been made taking $Cn$=0.015.

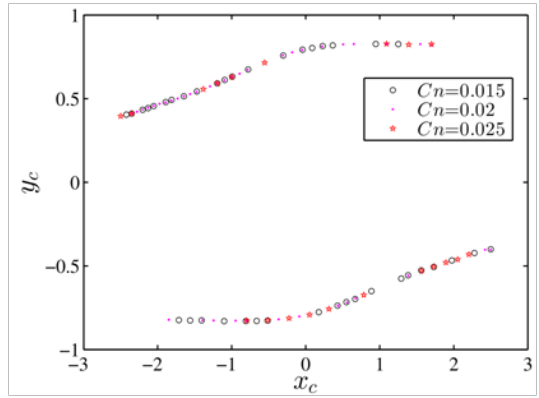

FIGURE 11. Grid independency check done for three different maximum element sizes (0.015, 0.02, 0.025) corresponding to $Ca = 0.2$, and 0.5. The other parameters are $S = 2$, $R$=0.5, $Re = 0.2$, $Wc$=0.2, $\lambda$= 1